\documentclass[epj]{svjour}
\usepackage{graphics}
\usepackage{float}
\usepackage{subfigure}
\usepackage{color}
\usepackage{geometry}
\usepackage{subfig}
\usepackage{caption}
\usepackage{amsfonts}
\usepackage{amsmath}
\usepackage{extarrows}
\usepackage{amssymb}
\usepackage{verbatim}
\usepackage{todonotes}
\usepackage{tikz}
\usepackage{slashed}
\usepackage{mathrsfs}
\usepackage{booktabs}
\usepackage{verbatim}
\usepackage{epsfig}
\usepackage{footnote}

\newcommand{\ket}[1]{|{#1}\rangle}

\newcommand{\bk}[1]{\langle{#1}|{#1}\rangle}

\newcommand{\bra}[1]{\langle{#1}|}
\newcommand{\bkt}[2]{\langle{#1}|{#2}\rangle}

\newcommand{\ot}{\otimes}

\newcommand{\av}[1]{\langle#1\rangle}

\begin{document}
\title{Amplification of rotation velocity using weak measurements in Sagnac's interferometer}
\author{Jing-Hui Huang\inst{1} \and Xue-Ying Duan\inst{2} \and Xiang-Yun Hu\inst{3}
\thanks{\emph{Present address:} jinghuihuang@cug.edu.cn}%
}                     
\offprints{}          
\institute{School of Institute of Geophysics and Geomatics, China University of Geosciences(Wuhan), China \and School of Automation, China University of Geosciences(Wuhan), China \and School of Institute of Geophysics and Geomatics, China University of Geosciences(Wuhan),China}

%
\date{Received: date / Revised version: date}
%
\abstract{
We study the amplification of rotation velocity with the Sagnac interferometer based on the concept of weak-value amplification. By using a different scheme to perform the Sagnac interferometer with the probe in momentum space, we have demonstrated the new weak measure protocol to detect the small rotation velocity by amplifying the phase shift of the Sagnac effect. At the given the maximum incident intensity of the initial spectrum, the detection limit of the intensity of the spectrometer and the accuracy of angular velocity measurement, we can theoretical give the appropriate potselection and the minimum of optical path area before experiment. In addition, we put forward a new optical design to increase the optical path area  and decrease the size of the interferometer to overcome the limit of instrument size. Finally, our modified Sagnac's interferometer based on weak measurement is innovative and efficient probing the small rotation velocity signal.
\PACS{
      {PACS-key}{discribing text of that key}   \and
      {PACS-key}{discribing text of that key}
     } 
} 
\maketitle
\section{Introduction}
\label{intro}
A general form of quantum weak measurement was first proposed by
Aharonov et al\cite{AAV},  where information is gained by weakly coupling the probe to the system. By appropriately selecting the initial and final state of the system, the measurement can be much larger than the eigenvalues of observables.  Thus, weak-values technique has been widely used to measure and amplify small physical quantities and effects which are not directly detected by conventional techniques in experiment, such as direct measuring the real and imaginary components of the wave-function \cite{2011Spectroscopy}, amplification of angular rotations \cite{PhysRevLett.112.200401}, longitudinal velocity shifts\cite{2013Weak}, frequency shifts\cite{PhysRevA.82.063822}, optical nonlinearities\cite{PhysRevLett.107.133603} and even the test of non-classical features in quantum  mechanics\cite{2015Weak,PhysRevLett.102.020404}. 

Although weak measurement have widely used to amplifying the small physical signal, its application in measuring rotation velocity was rarely reported. High-resolution rotation velocity measurements plays an important role in many emerging field. Especially in the rotational seismology, it has been noted by theoretical seismologists for decades that—in addition to translations and strains—the rotational part of ground motions should also be recorded\cite{Jacopo2012Horizontal,2009Introduction,0Broad}. The optical interferometer based on the Sagnac effect\cite{Heiner2007Broad} has made success in measuring rotation. 

\begin{figure}[htp!]
 \vspace{-0.2cm}
\centering
\resizebox{0.44\textwidth}{!}{%
  \includegraphics{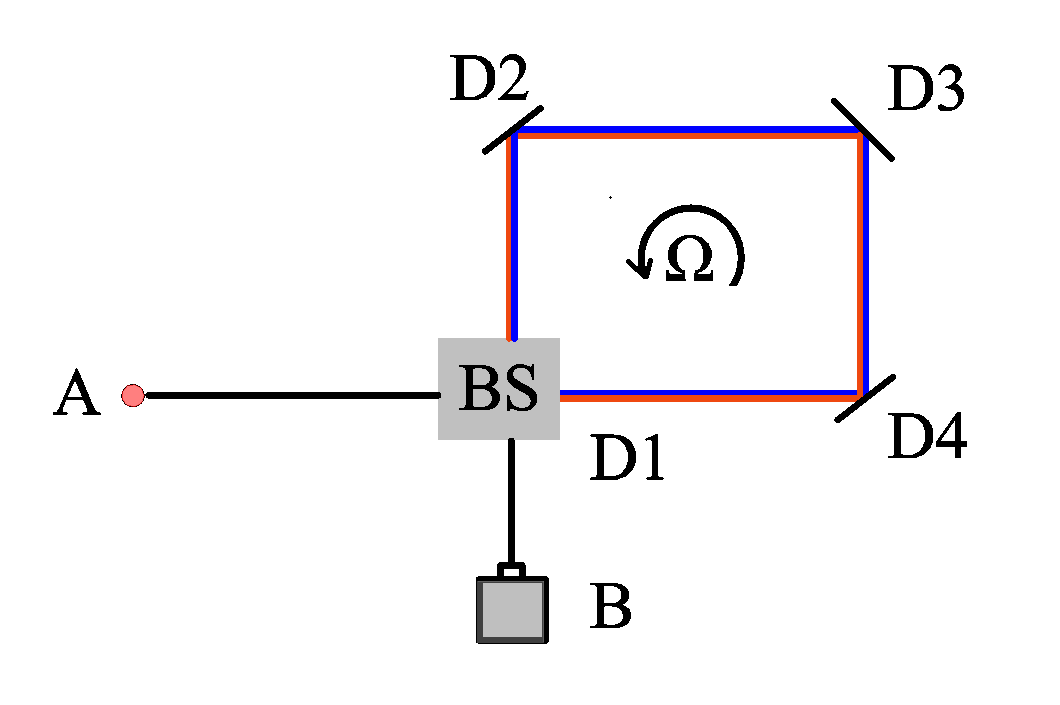}
}
\caption{Schematic of Sagnac's interferometer. A=light source; B= observer; BS = beam spliter ;  D1 and D2 are common mirrors; D3 is a special curved mirror }
\label{fig:sagnac_interferometer}       
\end{figure}
The basic principle of Sagnac's interferometer is given in Fig. \Ref{fig:sagnac_interferometer}. The light beam coming from the source A is split by the beam spliter C into a beam circulating the loop in clockwise D1D2D3D4D1 and a beam circulating the same loop in a counterclockwise direction D1D4D3D2D1. The two beams are reunited in B. When the whole interferometer is set in rotation with an angular rate of $\Omega$ rad/sec, a fringe shift $\Delta z$ with respect to the fringe position for the stationary is observed, which is given by the form
\begin{eqnarray}
\label{deltz}
\Delta z = \frac{4\Omega S}{\lambda_{0} c}
\end{eqnarray}
where S is the area enclosed by the light path. The vacuum wavelength is $\lambda_{0}$ and the free space velocity of light is c. Normally, in order to detecte the small rotation signal, the larger S is need. For example, the $4 \times 4 m$ ring laser system \cite{0Broad} that was installed at the geodetic observatory Wettzell, SE-Germany to detecting the earthquake-induced ground motions far from seismic sources. In the traditional  Sagnac's interferometer, the light intensity $I$ at B is recorded, which is dependence of the angular rate of $\Omega$ rad/sec:
\begin{eqnarray}
\label{deltII}
I=A[1+cos(2\pi \Delta z + \Delta \phi_{m})]
\end{eqnarray} 
Where $ \Delta \phi_{m}$ is the phase modulation to get the high sensitivity.  The outgoing light intensity $I$ can not be measured accurately and stably duo to the relative intensity noise of the light source, the loss and scattering of optical fibers, especially when the intensity of the detection light is very small. In that case, we put forward a modified Sagnac's interferometer based on weak measurement and analyze it in the momentum(frequency) domain. Our scheme is essentially different from the traditional Sagnac's interferometer, because the the angular rate of $\Omega$ do not depend on outgoing light intensity $I$, but is sensitive to the shifts of the center wavelength of the spectrum. As a result, we find that weak-value amplification of small rotation velocity can be performed with classical light with an finite bandwidth.\\
The rest of this paper is organized in the following way.
In Section 2, we give a brief review about weak measurements. In Section 3, we present a scheme for Amplification of rotation velocity using weak measurements in Sagnac's interferometer, and the numerical result is shown In Section 4. Finally, in Section 5, we give the conclusion about the work. Throughout this paper we adopt the unit $\hbar =1$.

\section{The review of Weak measurement}
\label{sec:deltz}
In this section, we briefly review the weak measurement proposed by Aharonov et al.~\cite{AAV}.
The weak measurement is characterized by state preparation, a weak perturbation, and postselection.
We prepare the initial state $\ket{\phi_{i}}$ of the system and $\ket{\psi_{i}}$ of the probe. 
After a certain interaction between the system and the probe, we postselect a system state $\ket{\phi_{f}}$ and obtain information about a physical quantity $\hat{A}$ from the probe wave function by the weak value
\begin{eqnarray}
\label{weak_value}
A_{w}:=\frac{\bra{\phi_{f}}\hat{A}\ket{\phi_{i}}}{\bkt{\phi_{f}}{\phi_{i}}},
\end{eqnarray}
which  can generally be a complex number.
More precisely, the shifts of the position and momentum in the probe wave function are given by the real and imaginary parts of the weak value $A_{w}$, respectively. 
We can easily see from Eq. \Ref{weak_value} that when $\ket{\phi_{i}}$ and $\ket{\phi_{f}}$ are almost orthogonal, the absolute value of the weak value can be arbitrarily large.
This leads to the weak-value amplification, as we will explain below.
As a trade-off, the probability of obtaining a postselected state that is almost orthogonal to the preselected state is very small. 
To make the large probe shift definite,  the weak measurement should be performed many times.

For the weak measurement, the coupling interaction is taken to be the standard von Neumann Hamiltonian, 
\begin{eqnarray}
H=g\delta(t-t_{0})\hat{A}\ot \hat{p},
\end{eqnarray}
where $g$ is a coupling constant and $\hat{p}$ is the probe momentum operator conjugate to the position operator $\hat{q}$. 
We have taken the interaction to be impulsive at time $t=t_{0}$ for simplicity. 
The time evolution operator becomes $\displaystyle e^{-ig\hat{A} \ot \hat{p}}$. 
After postselection, the probe state becomes 
\begin{eqnarray}
\ket{\psi_{f}}=\bra{\phi_{f}}e^{-ig\hat{A}\ot \hat{p}}\ket{\phi_{i}}\ket{\psi_{i}}.
\end{eqnarray}

We denote the expectation values of the initial and final probe momentum as
\begin{eqnarray}
\av{\hat{p}}_{i}:=\frac{\bra{\psi_{i}}\hat{p}\ket{\psi_{i}}}{\bk{\psi_{i}}}, \ \ \ \av{\hat{p}}_{f}:=\frac{\bra{\psi_{f}}\hat{p}\ket{\psi_{f}}}{\bk{\psi_{f}}}.
\end{eqnarray}
The shift of the expectation value of the momentum is defined by
\begin{eqnarray}
\label{qshifts}
\Delta\av{\hat{p}}:=\av{\hat{p}}_{f}-\av{\hat{p}}_{i}.
\end{eqnarray}

Here, we write  ${\Gamma}_{i}(p):=\bkt{p}{\psi_{i}}$ as  the initial probe wave functions in the  momentum spaces, respectively.

To see how the weak value emerges in theory, first consider the weak-coupling case following the original work~\cite{AAV}.
The probe state after the post selection becomes 
\begin{eqnarray}
\ket{\psi_{f}} &=&\bra{\phi_{f}}e^{-ig\hat{A}\ot \hat{p}}\ket{\phi_{i}}\ket{\psi_{i}} \nonumber \\
&=&\bra{\phi_{f}}\left[ 1-ig\hat{A}\ot \hat{p}\right]\ket{\phi_{i}}\ket{\psi_{i}}+O(g^{2}) \nonumber \\
&=&\bkt{\phi_{f}}{\phi_{i}}\left[ 1-igA_{w}\hat{p}\right]\ket{\psi_{i}}+O(g^{2}) \nonumber \\
&=&\bkt{\phi_{f}}{\phi_{i}}e^{-igA_{w}\hat{p}}\ket{\psi_{i}}+O(g^{2})
\end{eqnarray}
for $g |A_{w}|\ll1$.
The probe wave function in the position space after the postselection becomes
\begin{eqnarray}
\bkt{p}{\psi_{f}} &\approx& \bkt{\phi_{f}}{\phi_{i}}e^{-igA_{w}\hat{p}}\bkt{p}{\psi_{i}} 
\end{eqnarray}
and therefore, its absolute value squared $|\bkt{p}{\psi_{f}}|^{2}$ is obtained.
Thus, we obtain the shift of the expectation value of the probe momentum as 
\begin{eqnarray}
\label{p_linear}
\Delta\av{\hat{p}}= 2g W^{2} Im A_{w},
\end{eqnarray}
which is proportional to the imaginary part of the weak value. 

From Eqs. (\Ref{p_linear}), we can extract the weak value, and we can see that, as the weak value increases the probe position shift is amplified.
This effect is called the weak-value amplification. 
It is emphasized that the first-order approximation in $g$ and $|A_{w}|$ is used in the above calculation.

\section{Weak-value amplification on small rotation velocity}
In the following we will consider an weak measurement model probing the small ratation velocity,  meanwhile the numerical result and the discussion are shown. 
\subsection{Weak-value amplification on the Sagnac's interferometer}
\begin{figure}[htp!]
 \vspace{-0.2cm}
\centering
\resizebox{0.45\textwidth}{!}{%
  \includegraphics{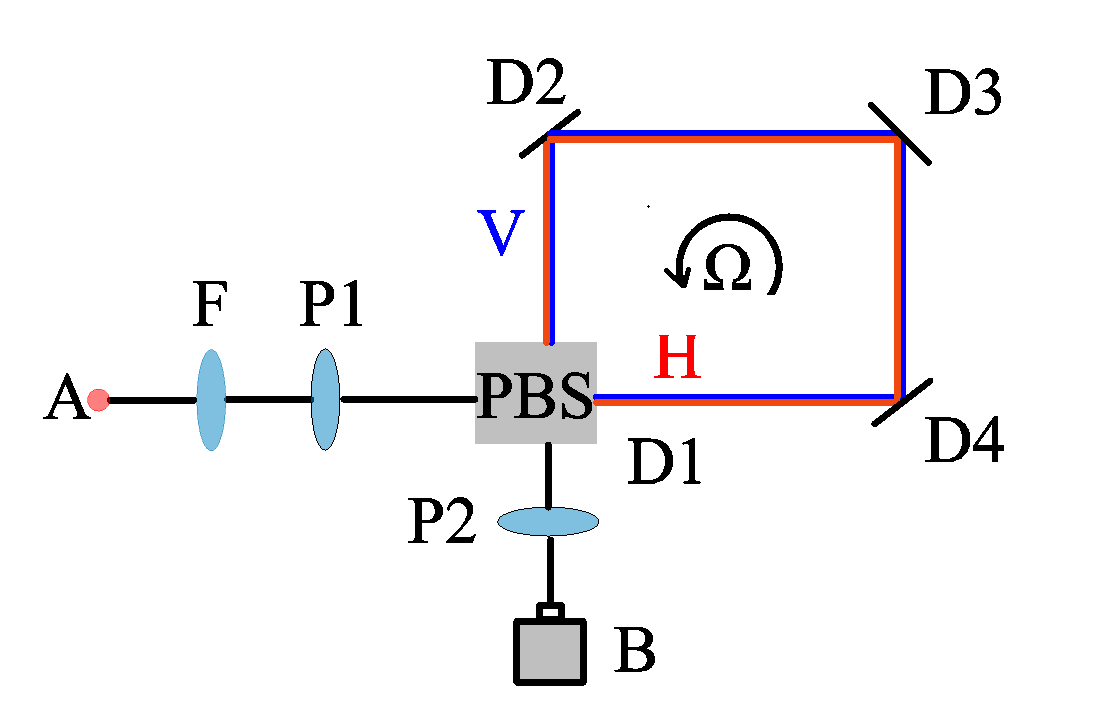}
}
\caption{Schematic of the modified Sagnac's interferometer based on weak measurements. A=light source; B= observer (the spectrometer) ; D1, D2, D3 and D4 are common mirrors; F is the Gaussian Filter; P1 is the polarizer corresponding with preselection,  P2 is the polarizer corresponding with postselection.}
\label{fig:interferometer}       
\end{figure}

We consider a small rotation velocity measurement on the Sagnac's interferometer. The Sagnac's interferometer we consider is shown in Fig.\Ref{fig:interferometer}. This scheme comprises there parts: state preparation, a weak perturbation, and postselection. 
Gwnerally, the state after preparation can be written as:
\begin{eqnarray}
\label{inter_sy_initial}
\ket{\phi_{i}}= sin(\alpha) \ket{H}+cos(\alpha) \ket{V} 
\end{eqnarray}

where $\ket{H}$ $\ket{V}$ represents the horizontal and vertical polarization, the angel $\alpha$ represents the angle between the direction of polarization $P1$ and the vertical direction.
The beam is injected into a Sagnac's interferometer, where the horizontally and vertically polarized components of the beam circulate in opposite directions.
Then the postselection stated can be obtained by adjusting the polarizer P2:
\begin{eqnarray}
\label{inter_sy_final}
\ket{\phi_{f}} =&-& cos(\alpha+ \beta) e^{-i \varphi /2} \ket{H}\nonumber \\
               &+&sin(\alpha+ \beta) e^{i \varphi /2}\ket{V}
\end{eqnarray}
where the angel $\beta$ represents the angle between the direction of polarization $P2$ and the horizontal direction of polarization $P1$, and the phase shift $\varphi$ is produced by  the Sagnac effect\cite{1967Sagnac} from Eq. (\Ref{deltz})
\begin{eqnarray}
\label{sagnac_affect}
\varphi=2\pi \Delta z =\frac{8\pi S}{\lambda_{0} c} \Omega
\end{eqnarray}
In our scheme, the observable $\hat{A}$ satisfies:
\begin{eqnarray}
\label{sy_observable}
\hat{A}= \ket{H} \bra{H}-\ket{V} \bra{V}
\end{eqnarray}
After weak correlation  and postselection, the probe wave function in the momentum space becomes 
\begin{eqnarray}
\bkt{p}{\psi_{f}} &=& \bkt{\phi_{f}}{\phi_{i}}e^{-igA_{w}\hat{p}}\bkt{p}{\psi_{i}} \nonumber \\
&=&m e^{-igp}\bkt{p}{\psi_{i}}+n  e^{igp}\bkt{p}{\psi_{i}}
\end{eqnarray}
with 
\begin{eqnarray}
\label{inter_weak_value1}
m=-sin(\alpha+ \beta) cos(\alpha) e^{i \varphi /2}\nonumber \\
n=cos(\alpha+ \beta) sin(\alpha) e^{-i \varphi /2}\nonumber
\end{eqnarray}
and therefore, its absolute value squared is given by:
\begin{eqnarray}
\label{eq_final_probe}
|\bkt{p}{\psi_{f}}|^{2} &=& |m+n|^{2}|cos(pg)+ImAw sin(pg)|^{2}\nonumber \\
&\times& |\bkt{p}{\psi_{i}}|^{2}
\end{eqnarray}
With the preselection and postselection, we can derive the so-called weak value(Eq. \Ref{weak_value}):
\begin{eqnarray}
\label{inter_weak_value}
A_{w}&=&\frac{\bra{\phi_{f}}\hat{A}\ket{\phi_{i}}}{\bkt{\phi_{f}}{\phi_{i}}}=\frac{m-n}{m+n}
\end{eqnarray}
$p=\frac{2 \pi}{\lambda}$
In our weak measurement protocol, we can get the shift of the center wavelength $\lambda_{0}$ from  the Eq. \Ref{p_linear} with $g=2 \pi / p_{0}$ \cite{2018Optical,Li:16}, $p_{0}$ corresponds to the center wavelength $\lambda_{0}=2 \pi / p_{0}$.
\begin{eqnarray}
\label{inter_delt_lambda}
\delta \lambda_{0}=-\frac{4 \pi (\Delta \lambda)^{2}}{\lambda_{0}} Im A_{w}
\end{eqnarray}

It is noteworthy that under actual experimental conditions, it is efficient and convenient to recording the spectrum and fitting the central wavelength of the spectrum with the Gauss function. Finally, the relationship of the shift of the central wavelength $\delta \lambda_{0}$ and the rotation velocity $\Omega$ of the interferometer can be obtain from Eq. (\Ref{inter_delt_lambda}), because in Eq. (\Ref{inter_delt_lambda})  the rotation velocity $\Omega$ is hidden variable of the weak value $A_{w}$. In addition, the sensitivity of our scheme can be defined:
\begin{eqnarray}
\label{inter_delt_semsiti}
k=abs(\frac{d \delta \lambda_{0}}{d \Omega})
\end{eqnarray}
which represents the ratio of the change in the output of the interferometer to the change in the corresponding input. In the absence of noise interference, greater sensitivity is our design goal.        
In this paper, the results of our simulation experiment will be shown in the next section.

\subsection{Numerical result and discussion}
\begin{figure*}[htp!]
	\centering
\subfigure
{
	\vspace{-0.2cm}
	\begin{minipage}{7cm}
	\centering
	\centerline{\includegraphics[scale=1.12,angle=0]{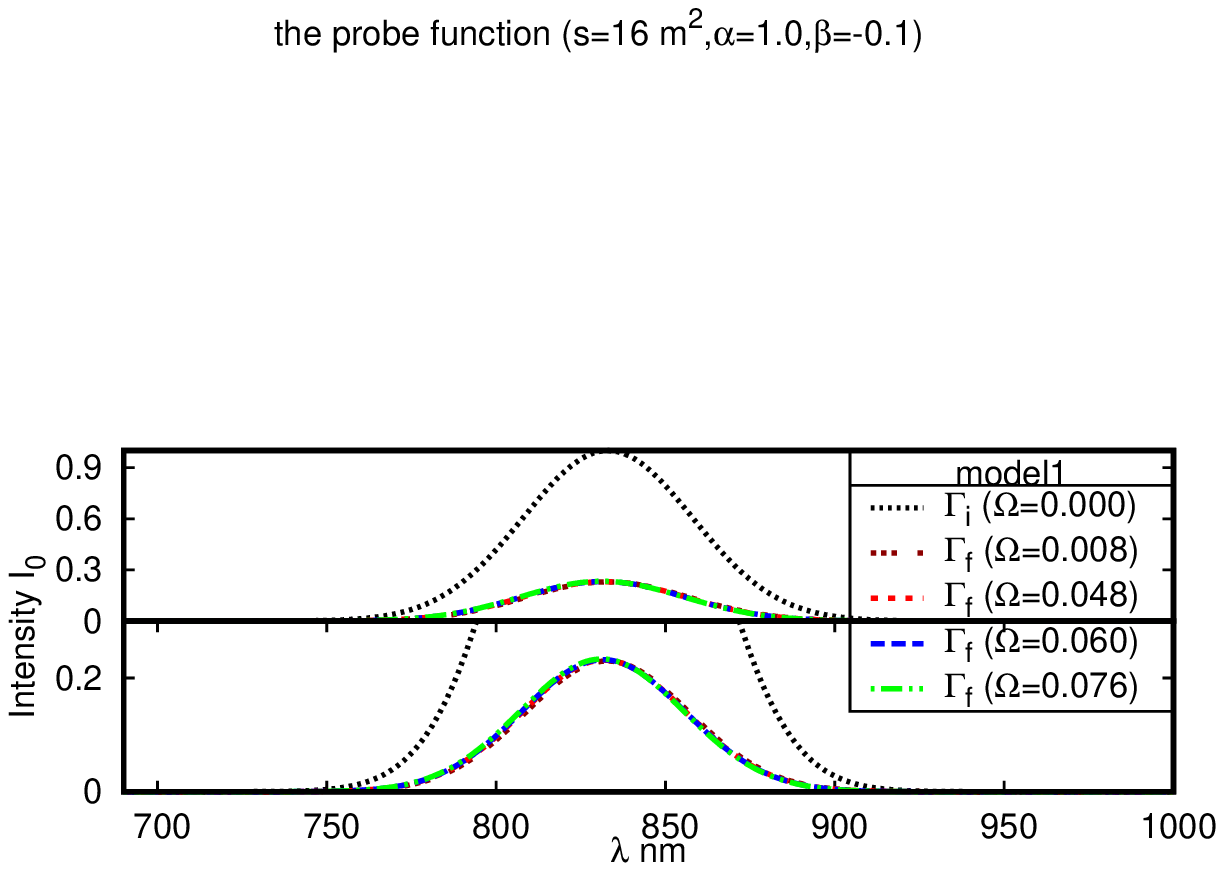}}
	\end{minipage}
}
\vspace{-0.2cm}

\subfigure
{
	\begin{minipage}{7cm}
	\centering
	\centerline{\includegraphics[scale=1.12,angle=0]{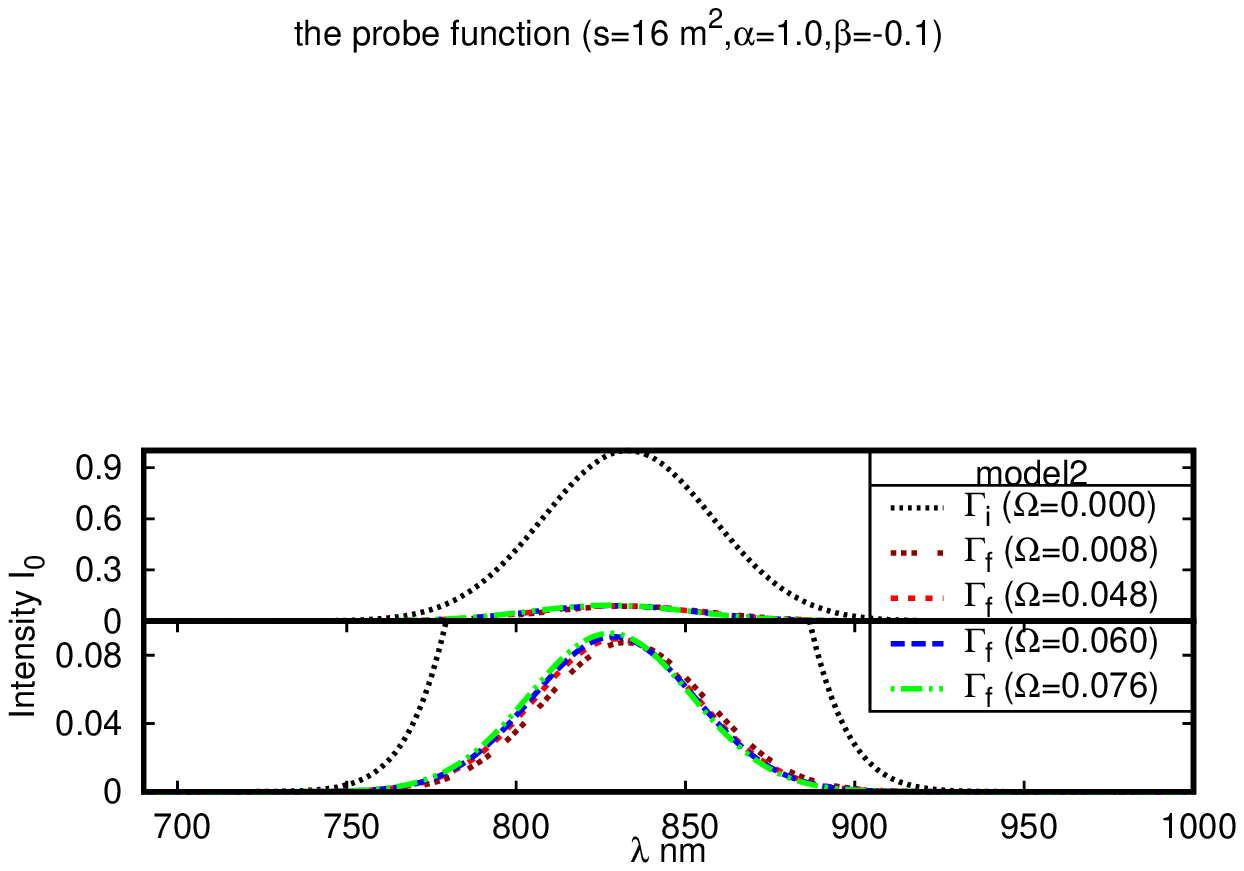}}
	\end{minipage}
}
\vspace{-0.2cm}

\subfigure
{
	\begin{minipage}{7cm}
	\centering
	\centerline{\includegraphics[scale=1.12,angle=0]{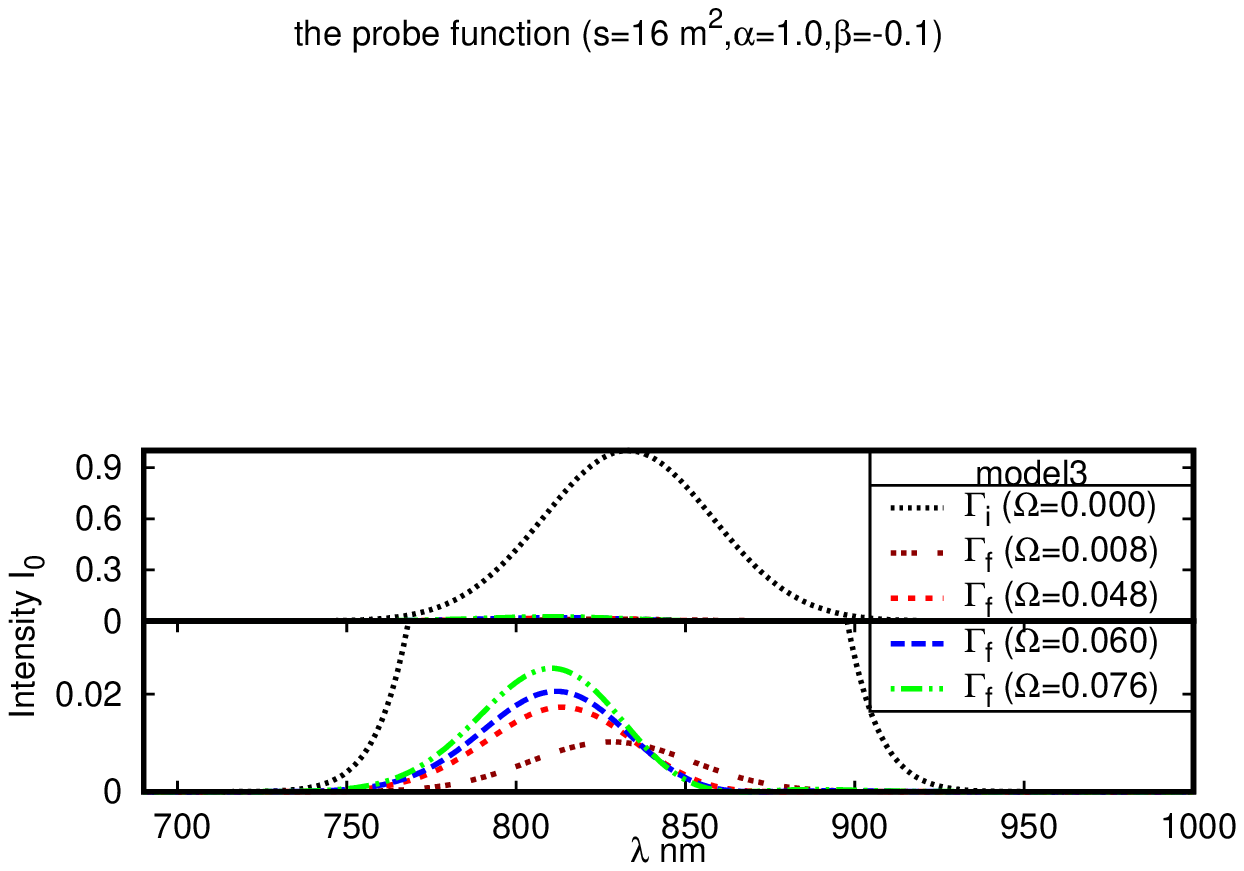}}
	\end{minipage}
}
\vspace{-0.2cm}

\subfigure
{
	\begin{minipage}{7cm}
	\centering
	\centerline{\includegraphics[scale=1.12,angle=0]{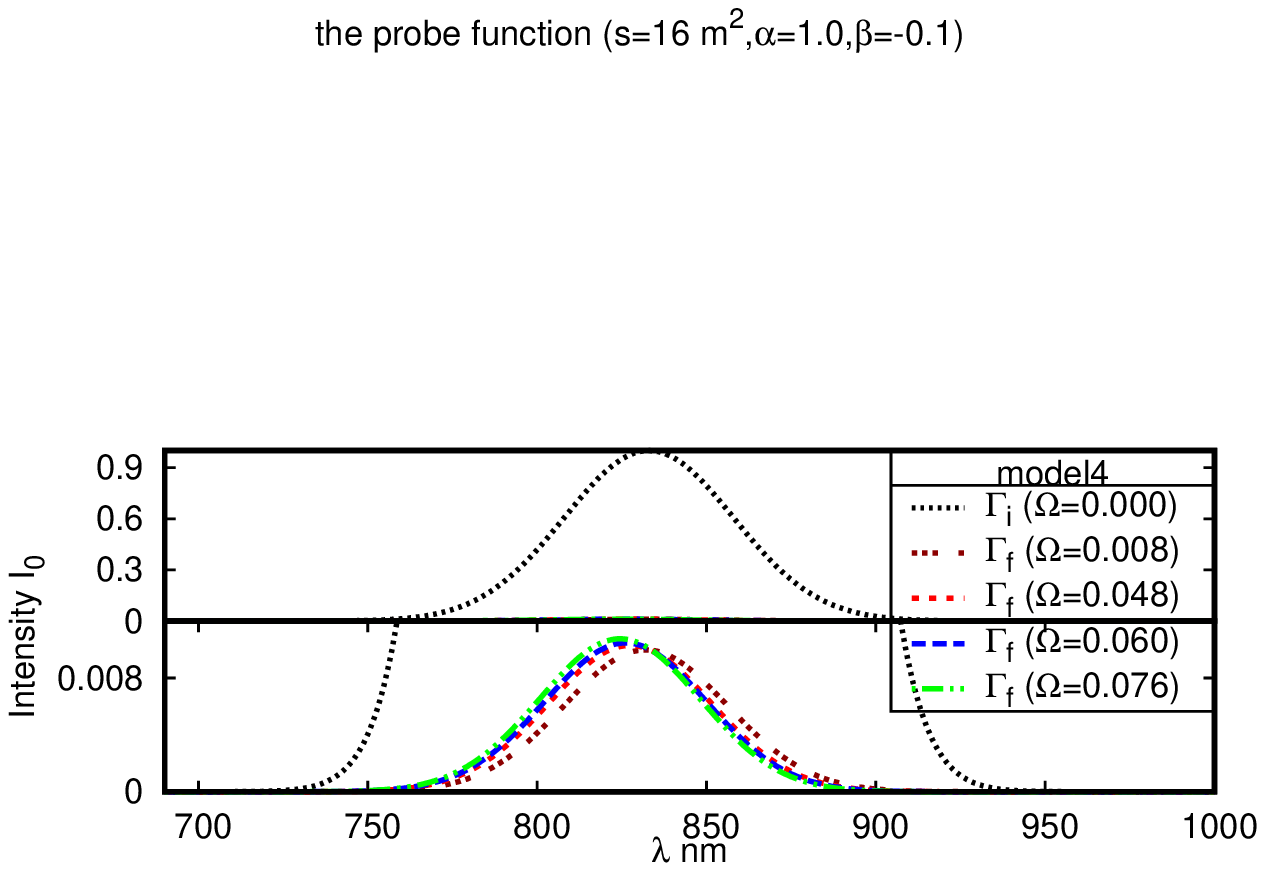}}
	\end{minipage}
}
\vspace{-0.2cm}

\vspace*{0mm} \caption{\label{result_numberical}The results of our simulation experiment with different model.  }
\end{figure*}

\begin{table}[htp!]
\centering
\caption{Parameters and the numerical values of our obtain results: S $[m^{2}]$ is the light area; $\alpha $ $[rad]$ is corresponding to the preslection and $\beta $ $[rad]$ is corresponding to the postslection; k [nm/(rad/sec)] is the sensitivity of our scheme and is the absolute value of the slope in the Fig. \Ref{result22}.}
\label{tab:1}       
\begin{tabular}{lllll}
\toprule
model & $S $ & $\alpha $ & $\beta $ & k\\
\noalign{\smallskip}\hline\noalign{\smallskip}
model1 & 16 & 0.1 & -0.5 & 9.9 $\times10^{1}$\\
model2 & 16 & 0.1 & -0.3 & 2.9 $\times10^{2}$\\
model3$^a$ & 16 & 0.1 & -0.1 & 1.9 $\times10^{3}$\\
model4 & 3 & 0.1 & -0.1 & 4.9 $\times10^{2}$\\
\toprule\\
\end{tabular}
\vspace*{0.1cm}  

\footnotesize{$^a$ The absolute value of the slope with model3 is calculated ( -0.03 $< \Omega < 0.03$ ) at small angular velocity.}
\end{table}

\begin{figure}[htp!]
 \vspace{-0.2cm}
\centering
\resizebox{0.45\textwidth}{!}{%
  \includegraphics{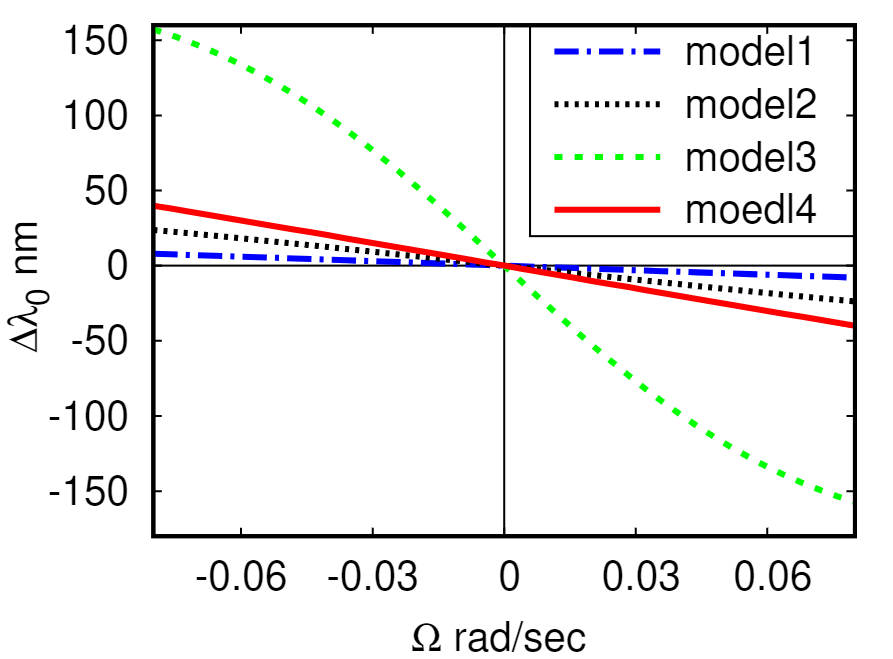}
}
\caption{The shift of the central wavelength $\delta \lambda_{0}$ dependence  of the rotation velocity $\Omega$ with different model.}
\label{result22}       
\end{figure}
In what follows, we fix a specific experimental setup, that is, the given chosen preselected state, postselected state and the initial probe in momentum space, so that we can calculate the weak value and the shift of the center wavelength of the probe before the experiment. In our numerical model, We take the initial probe wave function in 'Gaussian' form:
\begin{eqnarray}
\label{gaussian}
{\Gamma}_{i}(p)={\Gamma}_{i}(\lambda) =I_{0}e^{-(\lambda-\lambda_{0})^{2}/W^{2}},
\end{eqnarray}
where $W=\Delta \lambda$ is the variance of the initial probe, $\lambda_{0}$ is the center wavelength of the initial probe, $I_{0}$ is the maximum incident intensity of the initial spectrum. It is noted that the  wavelength and the momentum of the probe has the relationship $p=2 \pi / \lambda$. Then, by using the Eq. (\Ref{eq_final_probe}) and Eq. (\Ref{inter_delt_lambda}), we obtain probe function and the shifts of the center wavelength with different values of $S$,  and $\beta$ in Fig. (\Ref{result_numberical}) and Tab. (\Ref{tab:1}).

Fig. (\Ref{result_numberical}) with model1, model2 and model3 show the results of our simulation experiment with different value of $\beta$ when the optical path area $S=16 m^{2}$ and the angler $\alpha=0.1$ rad are given. The the optical path area $S=16 m^{2}$ is fixed when referring to the actual optical path area in the $4 \times 4 m$ ring laser system \cite{0Broad}. From the Eq. (\Ref{weak_value}), we could conclude that the more large amplification is set, the value of the $\bkt{\phi_{f}}{\phi_{i}}$ need be smaller. Namely, The smaller anger $\alpha$ is, the greater shift of the center $\delta \lambda_{0}$ wavelength will be obtained. Because of the quantum mechanics principle of measurement, if the preselection state $\phi_{i}$ and the postselection sate $\phi_{f}$ tend to be orthogonal, we can hardly detect the final spectrum of the probe. The amplification of the rotation velocity can not be infinite due to the detection limit of the intensity of the spectrometer.

Fig. (\Ref{result_numberical}) with model3 and model4 show the results of our simulation experiment with different value of the optical path area $S$ when $\beta=0.1 $ rad and the angler $\alpha=0.1 $ rad are given. The larger the area S is, the the amplification effect is more obvious. It is convenient and necessary to design the optical path area smaller in the laboratory. Meanwhile, the minimum of optical path area is confined by the detection limit of the intensity of the spectrometer.

\subsection{A new optical design to increasing the optical path area}

Our numerical results show that increasing the optical path area $S$  is the efficient way to improving the sensitivity of measurement of the rotation velocity. Due to the limit of size of the interferometer, it is necessary to design a optic model with the bigger optical path area $S$  and with the small size of the interferometer. In this paper, we use the radius $R_{s}$ to describe the size of the interferometer. 

\begin{figure}[htp!]
 \vspace{-0.2cm}
\centering
\resizebox{0.45\textwidth}{!}{%
  \includegraphics{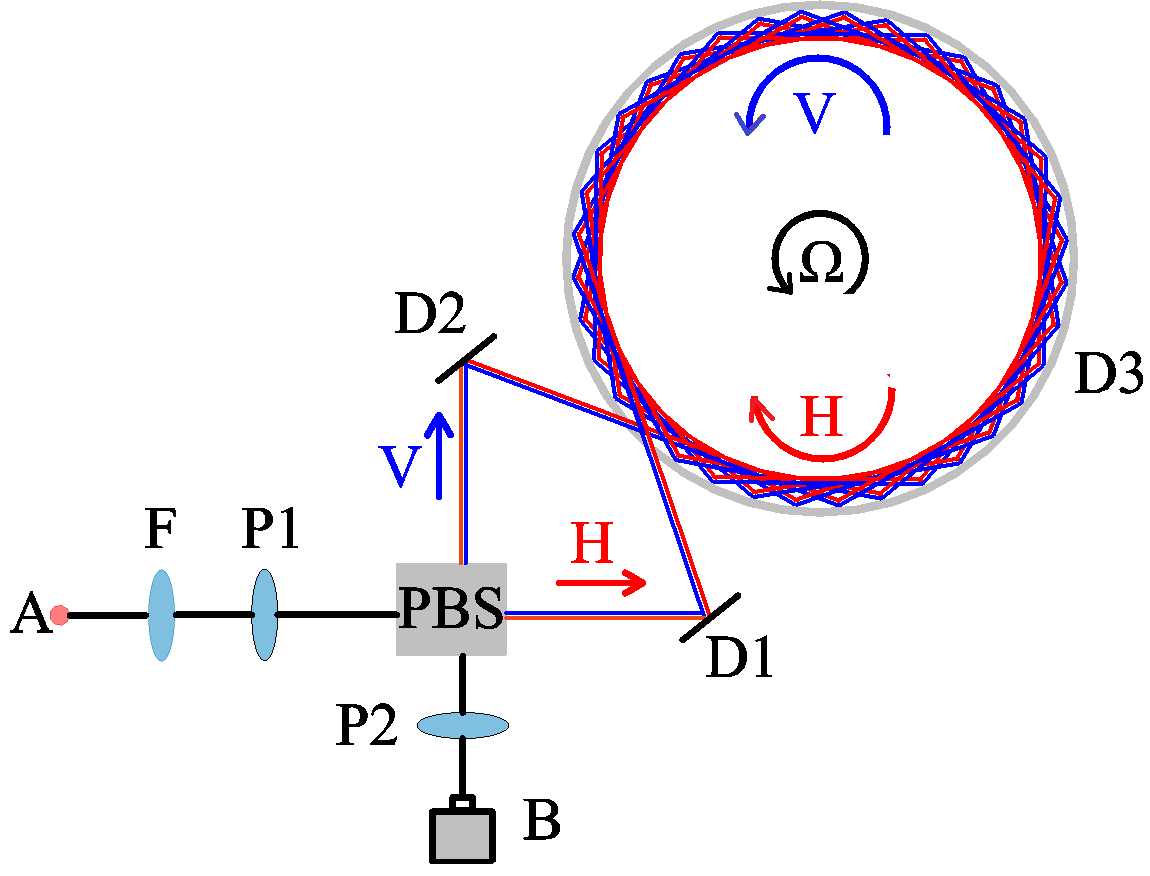}
}
\caption{The new optical design to increasing the optical path area. A=light source; B= observer (the spectrometer); F is the Gaussian Filter; D1 and D2 are common mirrors; D3 is a special curved mirror; P1 is the polarizer corresponding with preselection,  P2 is the polarizer corresponding with postselection.}
\label{fig:interferometer2}       
\end{figure}

In the Schematic of traditional Sagnac's interferometer(Fig. \Ref{fig:sagnac_interferometer}), the optical path can be replaced by N loops of fiber, so the fringe shift $\Delta z$ can be  be enlarged to $N \times \Delta z$. Usually, the fiber optic gyroscopes based on principle of traditional Sagnac's interferometer \cite{Yozo1996Fiber,articleChung} are commonly used for attitude and position control in various navigation systems. Due to the low loss in optical fibers, loops containing several kilometers of the fiber can be used in order to increase the sensitivity of the device. But in the modified  Sagnac’s interferometer(Fig. \Ref{fig:interferometer}), the scheme to replace the optical path with optical fiber seems inappropriate. Though the fiber can transfer polarized lights, it is difficult to technologically ensure that the polarization of the light at both ends of the fiber is consistent. In that case, we put forward a complex optical design to increasing theoptical path area, which is shown in Fig. \Ref{fig:interferometer2}.

\begin{figure}[htp!]
 \vspace{-0.2cm}
\centering
\resizebox{0.42\textwidth}{!}{%
  \includegraphics{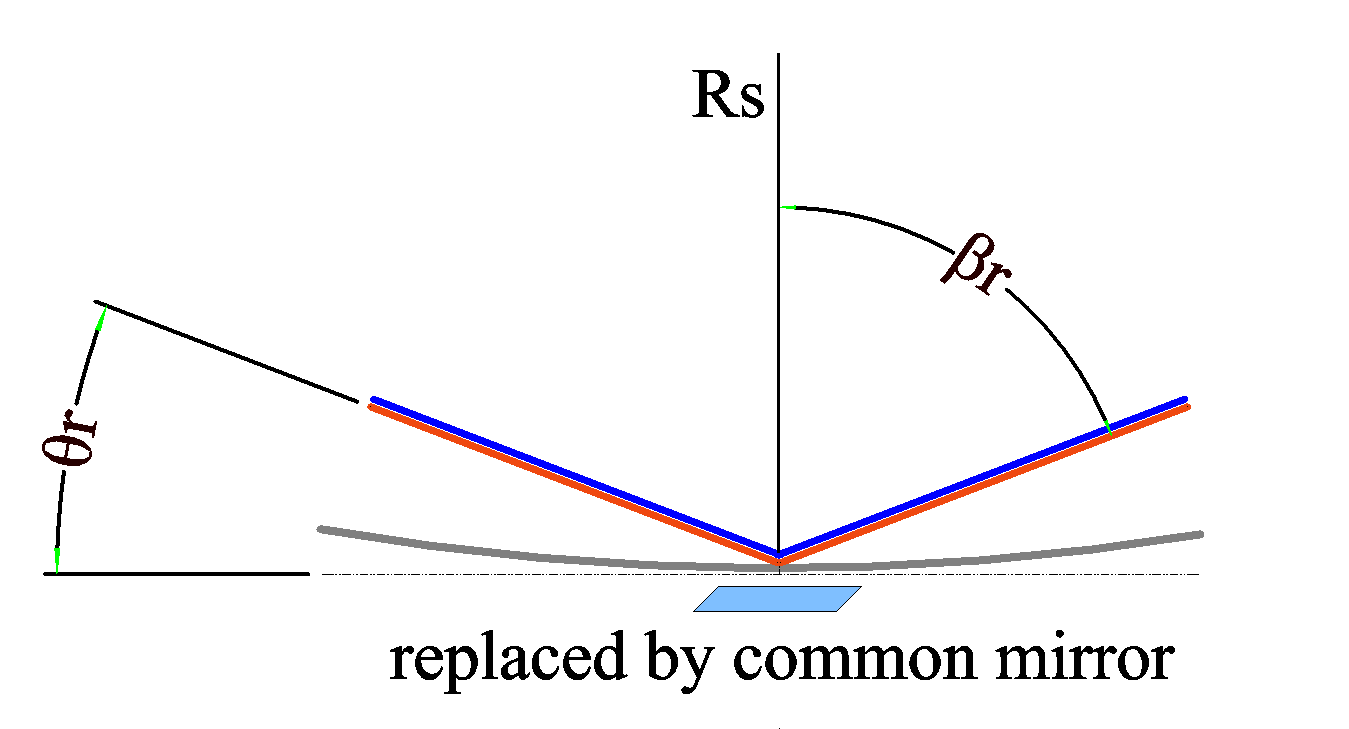}
}
\caption{The detail of the light passing through the interior of curved mirror D3. }
\end{figure}
The main difference between the scheme in Fig. \Ref{fig:interferometer} and the scheme in Fig. \Ref{fig:interferometer2} is how to realize the loop of the light. The new optical design is made to increase the optical path area $S$ and decrease the size $R_{s}$ of the interferometer. The special curved mirror D3 is made to ensure the light passing through the interiorof curved mirror D3 in multiple circles. Geometrical optics can decide the relationship of the number of turns ($N_{r}$) and the angle $\theta_{r} $ of incidence of light rays:
\begin{eqnarray}
\label{anew_1}
N_{r}=\frac{\mathcal{F}(2\theta_{r},360^{0})}{360^{0}}
\end{eqnarray}
where $\mathcal{F}(2\theta_{r},360^{0})$ is the function to obtain the least common multiple between $2\theta_{r}$ and $360^{0}$. Then we get the equivalent optical path area $S_{r}$:
\begin{eqnarray}
\label{anew_2}
S_{r}=\frac{\mathcal{F}(2\theta_{r},360^{0})}{2\theta_{r}} Rs^{2} sin(\beta_{r}) cos(\beta_{r})
\end{eqnarray}
where $\beta_{r}=90^{0}-\theta_{r}$. In the scheme(Fig. \Ref{fig:interferometer2}), we set $\theta_{r}=25^{0}$ and the equivalent optical path area $S_{r} \approx 14 Rs^{2}$ is obtained. When we comparing  $S_{r}$ to the path area $S=4 Rs^{2}$ of the scheme (Fig. \Ref{fig:interferometer}) with square length $2Rs$, the new design (Fig. \Ref{fig:interferometer2}) could increase the equivalent optical path area about 3 times the optical path area in the scheme (Fig. \Ref{fig:interferometer}) with the same size of the interferometer. Finaly, the above discussion shows the new optical design is efficient way to  increase the optical path area $S$ and decrease the size $R_{s}$ of the interferometer.

\section{Conclusion}
In conclusion, we use weak measurement to probe the small rotation velocity in Sagnac's interferometer. By choosing the appropriate preselection, postselection, the initial probe in the momentum space and the size of the Sagnac's interferometer, we obtain the relationship between the shifts of the center wavelength and the rotation velocity. This amplification effect can’t be explained by classical wave interference \cite{Heiner2007Broad,1991Optical}, due to the statistical feature of preselection and postselection with disturbance\cite{PhysRevLett.113.120404}. 

Our scheme is essentially different from the traditional Sagnac's interferometer, because the the angular rate of $\Omega$ do not depend on outgoing light intensity $I$, but is sensitive to the shifts of the center wavelength of the spectrum. At the given the maximum incident intensity $I_{0}$ of the initial spectrum, the detection limit of the intensity of the spectrometer and the accuracy of angular velocity measurement, we can theoretical give the appropriate potselection and the minimum of optical path area before experiment. In addition, we put forward a new optical design increase the optical path area $S$ and decrease the size $R_{s}$ of the interferometer. Finally, our modified Sagnac's interferometer based on weak measurement is efficient probing the small rotation velocity signal and the relevant optical experiments are taking in progress.
%
%
\section{Authors contributions}
All the authors were involved in the preparation of the manuscript.
All the authors have read and approved the final manuscript.

\section*{Acknowledgements}

We acknowledge financial support from National Natural Science Foundation of China (Grants No. 2018YFC1503700, No. G1323519204). 
%
%

\bibliographystyle{ws-mpla}
\bibliography{reference}

\end{document}